\documentstyle[12pt]{article}

\begin{document}

\author{A.P.~Potylitsyn \\
 Nuclear Physics Institute at Tomsk Polytechnic University,\\
 pr. Lenina 2A, Tomsk, 634050, Russia\\
 e-mail: pap@phtd.tpu.edu.ru}

\title{LASER POLARIZATION OF POSITRON BEAM.}

\maketitle
\date{}


 
\section{Introduction.}

  As it was pointed out in works \cite{1,2},
for a number of physical studies which are planned to be made with
the next generation colliders, it is necessary to use polarized
beams of both electrons and  positrons. The problem of producing
and acceleration of polarized electrons may be considered to be
solved \cite{3}, but the existing approaches to create polarized
positron beams [4--7] do not ensure the parameters required
(see \cite{8}).

This work proposes a new approach to produce polarized positron beams,
which is based on the process of multiple Compton backscattering of
unpolarized ultrarelativistic positrons in the field of  intense  
circularly polarized laser wave.

Under sufficiently high intensity of a laser flash a positron undergoes
 $N\gg$1 successive collisions with laser photons and at each collision
a positron loses a part of its initial	energy. A positron beam acquires
(on the average) a partial polarization along the direction coinciding
with the photon momentum.

As shown in work \cite{9} the deflection of a relativistic positron
 (electron) after a single act of Compton back-scattering (CBS) is
characterized by the angle  $\theta_e \sim
\displaystyle\frac{\omega_0}{mc^2}\ll \gamma^{-1}_0$,
if $\gamma_0 \leq 10^5$, here  $\omega_0$ is the laser photon energy 
 ($\sim $ 1 eV), $\gamma_0$
 is the Lorentz-factor of an initial particle. Therefore, one
can assume as a first approximation that the divergence of  positron
beam  does  not change, and, further it is possible to consider the
characteristics of a positron beam, as a  whole, after its
passage through a laser flash.

The quantization axis ($z$-axis) is chosen to be  parallel to the initial 
positron momentum (i.e., antiparallel to the laser photon momentum). 
It is clear that for obtaining a nonzeroth polarization of the final  beam, 
it is necessary that after a multiple process of CBS the "occupancies" of 
spin states should differ:
\begin{eqnarray}
 N_{\uparrow}\neq N_{\downarrow}\;,\nonumber
\end{eqnarray}
if for the initial unpolarized beam  
\begin{eqnarray}
 N_{\uparrow}= N_{\downarrow}\;. \nonumber
\end{eqnarray}
 
In a single  act of CBS the spin flip processes are described by
$w_{\uparrow \downarrow}$ and $w_{\downarrow \uparrow }$ probabilites.
In case when  $w_{\uparrow \downarrow} =
 w_{\downarrow \uparrow }$,  the process of laser polarization is
impossible, while at   $w_{\uparrow \downarrow}
\neq w_{\uparrow \downarrow} $ in the process of multiple  Compton
scatering the positron beam can achieve the polarization degree equal to:
\begin{eqnarray}
\xi_{z_{\scriptscriptstyle max}} = \frac{w_{\uparrow \downarrow}
- w_{\downarrow \uparrow }} {w_{\uparrow \downarrow}
+ w_{\downarrow \uparrow }}\;. \nonumber
\end{eqnarray}
Here $\xi_{z_{\scriptscriptstyle max}}$ is the mostly achievable
value of the $z$--component of the final beam  polarization.

In my previos work \cite{10} the wrong estimation of the final 
polarization degree (close to unit) was obtained. As was shown by G. L. Kotkin,
H. Perlt and V. G. Serbo \cite{13} positrons passing through an intense laser
field region change their polarization even without interaction with laser photons.
It leads to the decreasing of the attainable polarization value.
 
In the present work this effect was taken into account.
Also the expressions for  $\omega_{\uparrow \downarrow}$
   and	$\omega_{\downarrow \uparrow }$, were obtained and it was shown
that $\omega_{\uparrow \downarrow}\neq \omega_{\downarrow \uparrow }$;
the requirements to the parameters of the laser flash to obtain the significant
polarization  were analyzed; the process of laser polarization of particles
 in the storage ring during a short-time period was considered.

\section{Compton-effect cross-section on longitudinally 
 polarized electron (positron) with spin flip.}

The process of Compton backscattering (CBS) will be considered in a system
where an electron (positron) is at  rest (rest frame - RF), and a circularly
polarized  laser photon moves against the axis $z$.
We will be also interested in the component of the recoil electron spin
along the axis	$z$, since, because of the azimuthal symmetry of the
process after averaging over the assembly  of the initial beam particles,
the transverse projections of the spin are set to zero.

The authors of classical  work	\cite{11}
calculated the cross-section of Compton effect for an electron at rest,
considering  the correlations among polarizations of all four particles 
taking  part in the reaction.

After summing up over polarizations of	a scattered photon  according
to \cite{11}, one can get

\begin{equation} \label{l1}
 \frac{d\sigma}{d\Omega}=  2r^2_0 \Big(\frac{k}{k_0}\Big)^2
\Big\{\Phi_0 +P_c\xi_{0z}\Phi_1 +P_c\xi_{z}\Phi_2 +
\xi_{0z}\xi_{z}\Phi_3 \Big\}\;.  
\end{equation}

Here $r_o$ -- a classical radius of an electron;  $k_0$ and $k $ --
are the energies of the initial and scattered photons;
 $P_c$ -- is the  degree of a circular polarization of	a laser
photon; $\xi_{oz}, \xi_z$ -- are the spin components of initial
and scattered electrons;
the functions of $\Phi_i$	will be determined later.
 A similar cross-section for a positron may be obtained after
 a simple  changing the	signs of the
terms proportional to $P_c$.

It should be pointed out that in (\ref{l1}) the term which is 
proportional to $P_c \xi_{0z} \xi_z$ is absent.

For electrons with the initial energy  $\gamma_0 \le 10^4$
and the energy of the laser photon  $\omega_0 \sim $ 1 eV in the
laboratory frame (LF) the initial photon energy in the	RF system is
obtained after the Lorentz transformation
(the system  $\hbar = m=c=1$ is used)
\begin{equation}\label{l2}
 k_0 = (1+\beta_0)\gamma_0\omega_0 = 2\gamma_0\omega_0 << 1 \;,
\end{equation}
and the energy of a scattering photon may be obtained from the
conservation laws:
\begin{equation}\label{l3}
 k = \frac{k_0}{1+k_0(1 - \cos\theta)} \approx k_0 [1-k_0(1-\cos\theta)]\;.
\end{equation}
Here $\theta$ -- is the angle of a scattered photon  in the RF.

In equation (\ref{l3}) the terms of order of  $k_0^3$	and higher 
are neglected.
This approximation will be also used further. Let us write down the functions
 $\Phi_i(k_0, \cos\theta)$:
$$
\Phi_0 = \frac{1}{8}\Big[1+\cos^2\theta + k^2_0 (1-\cos^2\theta)\Big]\;,
$$
$$\Phi_1 = -\frac{1}{8} \cos\theta (1-\cos\theta)
[2k_0-k^2_0 (1-\cos\theta)]\;,
$$
\begin{equation}\label{l4}
\Phi_2 = - \frac{1}{8}(1-\cos\theta)[2k_0\cos\theta -
 k^2_0 (1+\cos\theta-2\cos^2\theta)]\;,
\end{equation}
$$
\Phi_3 = \frac{1}{8}\Big[1+\cos^2\theta - k^2_0 \Big(\frac{1}{2}+\cos\theta
- 2 \cos^2\theta + \cos^3\theta \Big) \Big]\;.
$$

From  (\ref{l1}) and (\ref{l4})  after simple integration one has:
$$
\sigma =\pi r^2_0 \Big[\frac{4}{3}\Big(1-2k_0 + \frac{26}{5}k^2_0\Big)
+ P_c\xi_{oz} \frac{1}{3}\Big(2k_0 - 10 k^2_0\Big) + P_c\xi_z \frac{1}{3}
\Big(2k_0 - 8k^2_0\Big) +
$$
\begin{equation}\label{l5}
+ \xi_{0z}\xi_z \frac{4}{3}\Big(1-2k_0 + \frac{22}{5}k^2_0\Big)\Big]=
\sigma_0+P_c\xi_{0z}\sigma_1 + P_c\xi_z\sigma_2 + \xi_{0z}\xi_z\sigma_3.
\end{equation}
Section (\ref{l5}) is invariant and  may be used both in the RF,
and the LF.
The probability of the Compton effect with a spin flip is proportional to
cross-section (\ref{l5}) when   $\xi_{0z}$ and $\xi_z$ have opposite signs:
$$
 w_{\uparrow \downarrow} = w(\xi_{0z}=+1, \ \xi_z = -1)=
\rm const \{\sigma_0 +P_c\sigma_1 - P_c\sigma_2 -\sigma_3\}
$$
$$
w_{\downarrow \uparrow }= w(\xi_{0z}= -1, \ \xi_z = +1)=
\rm const \{\sigma_0 -P_c\sigma_1 + P_c\sigma_2 -\sigma_3 \}
$$
As it follows from (\ref{l5}) $\sigma_1 \neq  \sigma_2$,
therefore, the polarization of scattering electrons in CBS
(laser polarization) is, in principle, possible.

Cross-section (\ref{l5}) can be derived from the invariant expressions 
with taking into account the polarization of all particles \cite{12,13}.
 Thus, the authors in \cite{13} obtained the CBS section
depending on the longitudinal polarization of a scattered
electron (i.e., on the projection of a scattered electron spin on its
momentum) in the invariant form. Following from the above work, let us
 write down the cross-section where the component of a scattered electron 
polarization is maintained along the same  axis $z$ as that of the 
initial electron:
\begin{equation}\label{l6}
\frac{d\sigma}{d y} = \pi r^2_0 \frac{1}{x}\Big\{F_0 + (s_z G_2 +
c_z G_3)\xi_z\Big\}
\end{equation}
 In (\ref{l6}) averaging over the azimutal angle was performed.
 The invariants  $x,y$ are found in a standard way:
\begin{equation}\label{l7}
x =2pk, \ y = 1 -\frac{pk^{'}}{pk}
\end{equation}

where $p, k, k^{'}$ are the four-vectors of the  initial  electron and photon
and the scattered   photon, respectively. The functions $F_0, G_2, G_3$
are determined as:
$$F_0=\frac{1}{1-y}+ 1-y -s^2 +y \frac{2-y}{1-y} \ c \xi_{0z}P_c
$$
\begin{equation}\label{l8}
G_2=	y s c P_c + (1+c^2- y c^2 )\ s \xi_{0z}\;,
\end{equation}
$$
G_3=\Big(\frac{y}{1-y}+yc^2\Big) P_c + \Big[\frac{1}{1-y}+
(1-y) c^2  \Big] \ c\xi_{0z}\;,
$$
where $s = 2\sqrt{r(1-r)}, \ c = 1-2r, \ r=\displaystyle\frac{y}{(1-y)x}$.

The coefficients $s_z,\  c_z$ are none other than the coefficients of the
rotation matrices from the basis used in \cite{12} to the basis 
where one of the  axes coincides with	the axis $z$.

Substituting   (\ref{l1}) into (\ref{l6}) let us write down the section 
in the form as in (\ref{l1}):

$$ 
\frac{d\sigma}{d y} = \pi r^2_0 \frac{1}{x} \Big\{ \frac{1}{1-y}
+1-y-s^2 + \xi_{0z}  P_c c y \ \frac{2-y}{1-y}+
 $$
$$
+P_c\xi_z \big[s_z y s c + c_z \Big(\frac{y}{1-y}+y c^2 \Big)\big]
+ \xi_{0z}\xi_z \Big[s_z s(1+c^2 -y c^2)+ 
$$
\begin{equation}\label{l9}
+ c_z c \Big(\frac{1}{1-y} + (1-y)c^2\Big)\Big]\Big\}
= \pi r^2_0 \frac{1}{x} \Big\{\tilde{\Phi_0}+ \tilde{\Phi_1}P_c\xi_{0z}+
  \tilde{\Phi_2}P_c \xi_{z} + \tilde{\Phi_3} \xi_{0z}\xi_z \Big\}
\end{equation}
One can easily see that, if  $s_z=s,\ c_z=c $,(which corresponds to the
longitudinal polarization of a scattered electron), 
 $\tilde \Phi_1=\tilde \Phi_2$.
However, generally speaking, $s_z\neq s,\ c_z\neq c$,
which can be easily seen in the RF. Let us  calculate section 
 (\ref{l9}) in the  RF and compare with the results obtained earlier
(Eqs. (\ref{l4}), (\ref{l5})).
Coefficients $s_z, c_z$ are defined as
\begin{equation}\label{l10}
 s_z = - \frac{\vec{k_0}}{k_0}\ {\vec{n}\hspace{2pt}'_2} \ \
c_z = - \frac{\vec{k_0}}{k_0} \ {\vec{n}\hspace{2pt}'_3}\;,
\end{equation}
 where the unit vectors $\vec{n}\hspace{2pt}'_{i}$ determine the basis used in
\cite{12}. In the  RF by neglecting the terms  $\sim k^3_0$
and higher, one has from  (\ref{l10})

$$
s_z = \frac{\sin\theta (1+k_0)}{1+k_0(1-\cos\theta)}\Big[1 - k^2_0
\frac{1-\cos\theta}{2}\Big]\;,
$$
\begin{equation}\label{l11}
c_z = \frac{\cos\theta - k_0(1-\cos\theta) - k^2_0(1-\cos\theta)\cos\theta}
{1+k_0(1-\cos\theta)}\;,
\end{equation}
while   in the same system (see \cite{13})
\begin{equation}\label{l12}
s =\sin\theta, \ c =\cos\theta\;.
\end{equation}
Passing from the invariant variables to those used earlier 

\begin{equation}\label{l13}
 x =2k_0, \ y = 1-\frac{k}{k_0} = \frac{k_0(1-\cos\theta)}
{1+k_0(1-cos\theta)}
\end{equation}
and substituting (\ref{l11}), (\ref{l12}), (\ref{l13}) into 
cross-section (\ref{l9}), one can show that the section  obtained 
completely coincides with  (\ref{l5}), in the same approximation as above.

\begin{center}
\section{Process of Multiple Compton Backscattering.
Main characteristics.}
\end{center}

The number of hard photons  $N$ produced in colliding  ofone electron with 
photons of the laser flash is a random value as well as the other 
characteristics of the  process of multiple CBS (total radiation losses,
 the electron  multiple scattering angle, etc.).

For simplicity, further we will consider only mathematical expectations
 of these random values, omitting, as a rule  the averaging symbol
$N=<N>$.
In addition, we will deal with collisions of single positron bunches with 
photons. The  $N$ value (according to  work \cite{9} it is the conversion 
coefficient) can be found by using the luminosity of the process:
$$
N = \frac{L \sigma}{N_{e^{+}}}.
$$
Here $N_{e^{+}} $ is the number of positrons in a bunch;
$\sigma $ is the cross-section of CBS process; $L$ is the luminosity:
\begin{equation}\label{l14}
 L = 2 c N_0 N_{e^+}\int dV\int dt f_{ph}(x, y,z + c t)
f_e(x,y,z -\beta ct)\;.
\end{equation}
In (\ref{l14}) $N_{0}$ is the number of photons in a laser flash;
$f_{ph}, f_e$  are the normalized distributions of photons and positrons 
in bunches.

For the analytical estimations we will consider monodirected positron and 
photon  beams with the  Gauss distribution  in the  transverse and 
longitudinal directions:
\begin{equation}\label{l15}
f_e= \frac{2}{(2\pi)^{\scriptscriptstyle{3/2}}\sigma^2_e l_e}\exp
\Big\{-\frac{r^2}{\sigma^2_e}-\frac{(z-\beta c t)^2}{2 l^2_e}
\Big\}\;,
\end{equation}
$$
f_{ph} = \frac{2}{(2\pi)^{\scriptscriptstyle{3/2}}\sigma^2_{ph}l_{ph}}
\exp \Big\{-\frac{r^2}{\sigma^2_{ph}}-\frac{(z + c t)^2}{2 l^2_{ph}}\Big \}\;,
$$
$$
r^2 = x^2 + y^2.
$$
In this case (the so-called case of short-length bunches [10])
the luminosity is calculated analytically
\begin{equation}\label{l16}
 L= N_e^{+} N_{0}\frac{2}{\pi (\sigma^2_e+\sigma^2_{ph})}
\end{equation}
and it does not depend on the bunch lengths $2l_e, 2l_{ph}$ 
(in other words, on the time  of interaction).
For $k_0\ll1$ we have:
\begin{equation}\label{l17}
 N =2 N_{0} \frac{\sigma}{\pi(\sigma^2_e+\sigma^2_{ph})} \approx
\frac{2A}{\omega_0}\cdot \frac{2\sigma_0 }{\pi (\sigma^2_e + 
\sigma^2_{ph})}\;,
\end{equation}
where $A$ is the laser flash energy,  $\omega_0$ is the laser
photon energy.  The cross-section summed up over the spin of a 
scaterred positron and averaged over the spin of initial 
 $(\sigma_1\ll\sigma_0)$ is substituted into (\ref{l17}).
If the beam after the "damping-ring" is used as an initial one, in this case 
 $\sigma^2_e\ll\sigma^2_{ph}$ and, therefore, 
\begin{equation}\label{l18}
 N = \frac{A}{\omega_0}\ \frac{16}{3} \Big(\frac{r_0}{\sigma_{ph}}\Big)^2
= 4\frac{A}{\omega_0} \frac{\sigma_{\scriptscriptstyle{T}}}
{\lambda Z_{\scriptscriptstyle{R}}} = 2n_{\scriptscriptstyle{L}}
\sigma_{\scriptscriptstyle{T}} l_{ph}\;.
\end{equation}
Here  $\lambda$ is the wavelength of the laser photon;
$Z_{\scriptscriptstyle{R}}$ is Rayleigh length; 
$n_{\scriptscriptstyle{L}}$ is the photon concentration in the laser flash;
$\sigma_{\scriptscriptstyle{T}}= \displaystyle\frac{8}{3}\pi r^2_0$; \
$2l_{ph}$ is the length of the laser pulse.

Let us evaluate the average positron energy after the process of multiple 
CBS. 
The average energy after the first collision  act is found  from the 
condition
$$
<\gamma_1>=\gamma_0 - <\omega_1>\;,
$$
where $<\omega_1>$ is the average energy of an emitted 
photon in the LF, which depends on the average characteristics of a photon
in the  RF:
$$
<\omega_1> = \gamma_0 (<k> - \beta_0 <k_{\parallel}>),$$
\begin{equation}\label{l19}
<k> = \int k \Big(\frac{k}{k_0}\Big)^2 \Phi_0 d \Omega \Big/ \int
\Big(\frac{k}{k_0}\Big)^2 \Phi_0 d \Omega	=  k_0(1-k_0)\;,
\end{equation}
$$
<k_{||}> = \int k \cos \theta \Big(\frac{k}{k_0}\Big)^2 \Phi_0 d \Omega \Big/
 \int \Big(\frac{k}{k_0}\Big)^2 \Phi_0 d \Omega  \approx \frac{6}{5} k^2_0\;.
$$
Thus, 
$$
<\gamma_1> = \gamma_1 \approx \gamma_0(1-k_0) =
\gamma_0 (1-2\gamma_0\omega_0)\;.
$$
Before the second scattering act a laser photon in the RF will  have a 
smaller energy
$$
<k_1> = k_1= 2\gamma_1\omega_0 = k_0(1-k_0)\;.
$$
In a similar way one can obtain the relation   (further, for the sake of
simplificaton the averaging symbol is  neglected again)
\begin{equation}\label{l20}
 \gamma_{i+1} = \gamma_i (1 - 2\omega_0 \gamma_i)\;. 
\end{equation}
From (\ref{l20}) we get the equation in the finite differences
$$
\Delta\gamma_i = \gamma_{i+1} - \gamma_i = 2 \omega_0 \gamma^2_i \;.
$$
If the number of scattering acts  $k\gg1$, then,
passing from the finite difference equation  to the  differential one 
 and solving the latter, we have: 
\begin{equation}\label{l21}
 \gamma_k = \frac{\gamma_0}{1+2\gamma_0\omega_0 k}=
\frac{\gamma_0}{1+2\mu} \;.
\end{equation}
By making use of (\ref{l21}) one can estimate the total radiation losses by 
each  positron after  $N$ collisions
\begin{equation}\label{l22}
\Delta\gamma = \gamma_0 - \gamma_{\scriptscriptstyle{N}} =
   \gamma_0 \frac{2\mu}{1+2\mu} =
2\gamma^2_0\omega_0 \frac{N}{1+2\gamma_0\omega_0 N}\;.
\end{equation}
V.~Telnov \cite{14} proposed to use the process of a multiple CBS
for cooling the head-on electron (positron) beam.
Let us estimate the laser flash parameters and the number of hard photons
to decrease an electron initial energy by  one order of magnitude:
$$
\gamma_{\scriptscriptstyle{N}} = \frac{\gamma_0}{1 + 2\gamma_0\omega_0 N}=
\frac{\gamma_0}{10}\;.
$$
Hence, we  have the following relations for the collision number
$$N_{10}= \frac{9}{2\gamma_0\omega_0}\;,
$$
and laser energy:
\begin{equation}\label{23}
 A = \frac{9}{8} \frac{\lambda Z_{\scriptstyle{R}}}
{ \gamma_0  \sigma_{\tau}} \;.
\end{equation}

\section{Single-pass laser polarization of positron beam.}

If the initial positron is unpolarized and the laser photons possess
a 100 $\%$  right-hand circular polarization  ($P_c=+1$),
then, after the first scattering act the recoil positrons, on the whole,
become partly polarized. The value of the polarization component along the
$z$ axis can be found from cross-section (\ref{l5})
\begin{equation}\label{l24}
\xi_1 = \frac{\sigma_2 }{\sigma_0} \approx \frac{k_0}{2}(1-2k_0)=
\gamma_0\omega_0 (1-4\gamma_0 \omega_0)\;.
\end{equation}
Henceforth  the index  $z$  will be  neglected.
Taking into account that before  the second scattering act the 
photon energy in the RF equal to  $k_1=k_0(1-k_0)$
from (\ref{l5}) one can find the positron polarization after the second
scattering act:
$$ \xi_2 = \frac{\sigma_2 + \xi_1\sigma_3}
{\sigma_0 + \xi_1 \sigma_1} \approx
\frac{\frac{2}{3}k_1(1-4k_1)+ \xi_1 \frac{4}{3}(1-2k_1)}
{\frac{4}{3}(1-2k_1)+\xi_1 \frac{2k_1}{3}(1-5k_1)} \approx
$$
\begin{equation}\label{l25}
\approx k_0 (1-\frac{7}{2}k_0) = 2\gamma_0\omega_0 (1-7\gamma_0\omega_0)\;.
\end{equation}
Comparing (\ref{l25}) and (\ref{l24}) one can see that  $ |\xi_2|>|\xi_1|$.
By applying the method similar to the one used to derive (\ref{l21}) the
polarization value after $N$  collisions can be found
\begin{equation}\label{l26}
\xi_{\scriptscriptstyle N} = \frac{\gamma_0\omega_0 N}
{1+\gamma_0\omega_0 N} = \frac{\mu}{1+ \mu}\;.
\end{equation}
However, it should  be pointed out that  the immediate use of the
derived formula in order to evaluate the final polarization of the
positron beam is illegitimate. The problem is that, as a unpolarized 
positron beam is passing through a laser flash, the positrons which 
have  not  interacted with the laser photons become  polarized 
but in the opposite direction \cite{15}. 
Thus, the positrons in this part the beam will
have a nonzero polarization before the first collision act,
and that is why the use of formulas (\ref{l25},\ref{l26}) in this 
case gives  a wrong  results.

To obtain a correct result it is necessary to perform a Monte Carlo
 simulation of CBS process.
But the equilibrium polarization value can be found from the elementary
balance equations. In passing polarized positrons through a laser
flash having the photon concentration  $n_{\scriptscriptstyle L}$ 
hard photons are emitted in the process with a spin flip 
($\xi_z=-\xi_{0z}$), and without one ($\xi_z=+\xi_{0z}$).
The change of the number of positrons having the spins oriented along the
positive $N^{\uparrow}(\xi_z=+1)$  and negative direction of the 
quantization axis $N_{\downarrow}(\xi_z=-1)$ 
is described by the equation resulting from cross-section (\ref{l9}):
$$
 \frac{dN^{\uparrow}}{dz} = 2 n_{\scriptscriptstyle{L}}
[- (\sigma_0 + \sigma_1 - \sigma_2 -
\sigma_3) N^{\uparrow} + (\sigma_0 -\sigma_1 + \sigma_2 -\sigma_3)
N^{\downarrow}]\;,
$$
\begin{equation}\label{l27}
\frac{dN^{\downarrow}}{dz} = 2 n_{\scriptscriptstyle L}
[- (\sigma_0 + \sigma_1 - \sigma_2 -
\sigma_3) N^{\downarrow} + (\sigma_0 -\sigma_1 + \sigma_2 -\sigma_3)
 N^{\uparrow}]\;.
\end{equation}
By adding and subtracting Eqs. (\ref{l27})  we obtain standard balance 
equations \cite{16}:
$$\frac{d( N^{\uparrow} + N^{\downarrow})}{dz} =0
$$
\begin{equation}\label{l28}
\frac{d( N^{\uparrow} - N^{\downarrow} )}{dz} =  2n_{\scriptscriptstyle L}
[-2 (\sigma_0 -
\sigma_3)(N^{\uparrow} - N^{\downarrow}) - 2 (\sigma_1 -\sigma_2)
(N^{\uparrow} + N^{\downarrow})]\;. 
\end{equation}
Balance equations (\ref{l28}) have the  following  solution: 
\begin{equation}\label{l29}
\xi_z(z)= \frac{N^{\uparrow} - N^{\downarrow}}{N^{\uparrow}
+ N^{\downarrow}} = -
\frac{\sigma_1 -\sigma_2}{\sigma_0 -\sigma_3}\{{1- \exp [- 4 n_L
(\sigma_0 -\sigma_3)z]}\}\;.
\end{equation}
Here $z$ is the thickness of the "laser" target.
It follows from (\ref{l29}) that the maximum polarization degree 
(at $z\rightarrow + \infty$):
\begin{equation}\label{l30}
\xi_{z_{max}} = - \frac{\sigma_1 -\sigma_2}{\sigma_0 - \sigma_3}=
- \frac{5}{8} \;.
\end{equation}
The characteristic length of the laser bunch $Z_{pol}$,	
after passing which the positron beam acquires the polarization
degree ($1-e^{-1}) \xi_{max} \approx \  0.63 \ \xi_{max}$, 
is found from the following relation:
\begin{equation}\label{l31}
 Z_{pol} = \frac{1}{4 n_{\scriptscriptstyle L}(\sigma_0 -\sigma_3)}\;.
\end{equation}
From (\ref{l18}) and (\ref{l31})  we can find another characteristic 
of the laser polarization process -- the average number of $N_{pol}$
photons emitted by each positron after passing the length $Z_{pol}$:
\begin{equation}\label{l32}
 N_{pol}= \frac{8\pi}{3}n_{\scriptscriptstyle L} r^2_0 Z_{pol}
 = \frac{2 \pi r^2_0}{3} \
\frac{1}{\sigma_0 -\sigma_3}= \frac{5}{8 k^2_0} =
\frac{5}{32 \gamma^2_0 \omega^2_0}\;.
\end{equation}
Formula (\ref{l32}) is valid for  $k_0 = 2\gamma_0\omega_0 \ll 1$,
since it is in this approximation that the expressions for 
$\sigma_0 $ and  $\sigma_3$ (see (\ref{l9})) were obtained. 
The value  $N_{pol}$ does not depend on the specific laser parameters and
is a more suitable value, for example, in the case of a multiple passage 
of positrons through the laser bunch in the storage ring \cite{8}.

As it follows from (\ref{l32}) and (\ref{l18}) the characteristic number of 
hard photons necessary for cooling and polarization processes  depends on the 
energy of a laser photon in the RF such as $\sim 1/ k_0$ and $\sim 1 / k^2_0$.

But for the above case  $\gamma_0\omega_0 = 0.05$ and, consequently,
$$
N_{pol} =\frac{5}{32 (0.05)^2} = 62 \sim N_{10}\;.
$$

Having rewritten formula (\ref{l29}) in which one should pass from the laser  
bunch length  to the number of collision  acts (i.e., to the number of 
emitted photons (see (\ref{l18})):
$$ \exp(- \frac{z}{Z_{pol}})= \exp  (- \frac{n_{\scriptscriptstyle L}
 \sigma_{\scriptscriptstyle T} z}
{N_{pol}})= \exp(-\frac{N}{N_{pol}})\;,
$$
one gets the expression for the highest attainable polarization of the 
positron beam  

\begin{equation}\label{l33}
\xi_z = \xi_{z_{max}}  \Big[1- \exp \Big(-\frac{N}{N_{pol}}\Big)\Big].
\end{equation}
For the example considered  $\xi_z = 0.41$.

\section{Laser polarization in storage ring.}

It was shown in the previous part that by applying a sufficiently 
powerful laser one can, in principle, attain a polarized positron 
beam during a single pass through a laser flash. However, the laser 
parameters to create the significant  beam polarization with the 
time structure necessary 
for linear colliders are  beyond   the present-day 
technology.
An obvious way to increase luminosity of the process is to use a high 
Q-factor  optical resonator matched with the storage ring where a positron
beam is circulating.

In work \cite{17} the authors discussed the possibility of applying a 
laser with circularly polarized radiation in order to reduce the time of 
electrons self-polarization in the storage ring, and they showed that the CBS
process can be used for this purpose.

In his recent work \cite{8} J.~Clendenin proposed using a resonator and a
damping ring for simultaneous laser  cooling of a positron beam and its 
laser polarization. It is  clear that the requirements to the 
laser power are, in this case, considerably decreased 
(see the estimations the work cited).

J.~Clendenin considered a damping ring with a circumference of 297 m
for positrons with  $E$ = 1.98 GeV,
which is designed for synchrotron  cooling radiation.
For laser cooling of 100 MeV electrons the authors of  \cite{18}
proposed a small-sized ring  which   2 m in diameter.  It is of special 
interest to estimate the feasibility of this ring for  laser polarisation.

For a laser with the wavelength  $\lambda \sim$ 1 mcm ($\omega_0$ - 1.25 eV)
the parameter $k_0=2\gamma_0\omega_0 \approx 10^{-3}$, therefore,
$N_{pol}= 6.1 \cdot 10^5$.

During a single pass through the laser pulse having the energy 
$\sim$ 1J  and  $z_{\scriptscriptstyle{R}}$ = 1 mm each positron
experiences, on the  average, 
$N= \frac{2A}{\omega_0}  \frac{\sigma_{\scriptscriptstyle{T}}}{\lambda z_{\scriptscriptstyle{R}}} =0.82
$
collisions with laser photons, and the average energy of each emitted
photon  $<\omega_1> = 2\gamma^2_0\omega_0$.

Consequently, the energy losses by each positron  
 $\Delta\gamma =
4\sigma_{\scriptscriptstyle{T}} \gamma^2_0 {A} /
{\lambda z_{\scriptscriptstyle{R}}}$,
which coincides with formula (\ref{l3}) of work \cite{18}.

Thus, the necessary number of turns in the ring to attain 
$N_{pol}$ is:
$$n = \frac{N_{pol}}{N} = 7.4\cdot 10^5,
$$
which corresponds to the time  $\tau_{pol} ={\pi d}/{c}=
15.5 \cdot 10^{-3}$ sec. It seems that a serious problem   for such 
a ring is to maintain power in the resonator during the time mentioned.

It takes  $\sim 10^3$ laser injections into the resonator 
during the time $\tau_{pol}$ if  the cavity Q-factor $\sim 10^3$.

\section{Conclusion}

All the results were obtained with neglect of the contribution of the 
nonlinear processes. It is clear that for the laser flash parameters 
discussed in Parts 3 and 4 a linear approximation will be valid, if
the length of the laser pulse is rather high:
   $l_{ph} >>Z_{\scriptscriptstyle{R}}$.
To maintain a transverse size of the laser beam at a suitable level
along the full length  $l_{ph}$, it is necessary to apply a 
channeled laser beam, e.g., in a plasma channel
 \cite{19} or in a capillary  dielectric tube \cite{20}.

For a substantially nonlinear process, when during 
a single collision  act  $n>1$ laser photons are absorbed, the average energy
of  an emitted photon will considerably exceed the one for a linear process,
which will result in decrease in the value  $N_{pol}$. 
In other side it should be noted the value of polarization degree 
$\xi_{max} = 5/8$ was obtained in work \cite{21} for moving electron
in a long weak helical undulator (see, also, \cite{17}).

Authors of cited paper considered also the case of strong
helical undulator  where electron polarization may be much higher
($\xi_{max}$= 0.92). It is known the trajectory of electron in a helical
undulator is similar to one in a field of the intense circularly-polarized wave.
Having in mind the analogy between radiative processes in the helical
undulator and  circularly-polarized wave one may expect
that positron polarization during nonlinear multiple CBS process may exceed 
the value of (\ref{l30}) and approach to 92$\%$, typical  value for 
self-polarization mechanism due to synchrotron radiation.

It is  evidently the polarization of positron through nonlinear multiple CBS 
process demands the detailed investigations.

Author is grateful to V.~Telnov for stimulating criticism and to 
V.~Katkov, \\ V.~Strakhovenko, M.~Galynskii, E.~Bessonov and 
J.~Clendenin for useful discussions.

\end{document}